\documentclass[aps,twocolumn,showpacs,superscriptaddress,amsmath,amssymb,longbibliography]{revtex4-1}
\usepackage{graphicx}
\usepackage{color,soul}

\begin{document}

\title{Heterogeneous impact of a lockdown on inter-municipality mobility}

\author{Hygor P. M. Melo} \email{hpmelo@fc.ul.pt}
\affiliation{Centro de F\'{i}sica Te\'{o}rica e Computacional, Faculdade de
Ci\^encias, Universidade de Lisboa, 1749--016
Lisboa, Portugal}
  
\author{Jo\~ao Henriques} 
\affiliation{NOS, SGPS, S.A., 1600--404 Lisboa, Portugal}
\affiliation{Altran Portugal, 1990--096 Lisboa, Portugal}

\author{Raquel Carvalho} 
\affiliation{NOS, SGPS, S.A., 1600--404 Lisboa, Portugal}

\author{Trivik Verma} 
\affiliation{Faculty of Technology, Policy and Management,
Delft University of Technology, 2628BX Delft, the Netherlands}

\author{Jo\~ao P. da Cruz} 
\affiliation{Centro de F\'{i}sica Te\'{o}rica e Computacional, Faculdade de
Ci\^encias, Universidade de Lisboa, 1749--016
Lisboa, Portugal}
\affiliation{Closer Consultoria, 1070--101 Lisboa, Portugal}
\affiliation{Departamento de F\'{\i}sica, Faculdade de Ci\^{e}ncias,
Universidade de Lisboa, 1749--016 Lisboa, Portugal}

\author{N. A. M. Ara\'ujo} \email{nmaraujo@fc.ul.pt}
\affiliation{Centro de F\'{i}sica Te\'{o}rica e Computacional, Faculdade de
Ci\^encias, Universidade de Lisboa, 1749--016
Lisboa, Portugal}
\affiliation{Departamento de F\'{\i}sica, Faculdade de Ci\^{e}ncias,
Universidade de Lisboa, 1749--016 Lisboa, Portugal} 

\begin{abstract}
Without a vaccine, the fight against the spreading of the coronavirus has
focused on maintaining physical distance. To study the impact of such measures
on inter-municipality traffic, we analyze a mobile dataset with the daily flow
of people in Portugal in March and April 2020. We find that the reduction in
inter-municipality traffic depends strongly on its initial outflow. In
municipalities where the mobility is low, the outflow reduced by $10-20\%$ and
this decrease was independent of the population size.  Whereas, for
municipalities of high mobility, the reduction was a monotonic increasing
function of the population size and it even exceeded $60\%$ for the largest
municipalities. As a consequence of such heterogeneities, there were
significant structural changes on the most probable paths for the spreading of
the virus, which must be considered when modeling the impact of control
measures.
\end{abstract}

\maketitle

\begin{figure*}
\includegraphics[width=\textwidth]{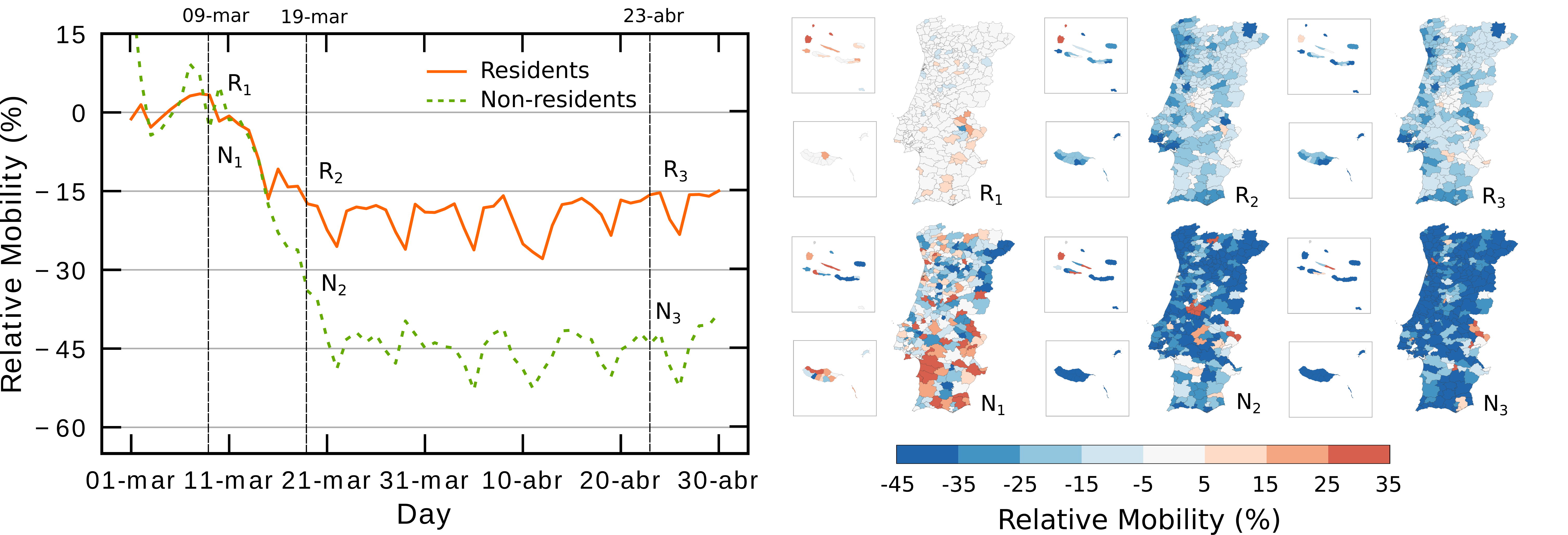}
\caption{\textbf{Effect of the lockdown in the mobility pattern of residents
and non--residents in Portugal.}  The relative mobility is calculated as the
total number of people that moved between two different municipalities on that
day divided by the average between March 2 and 6 (see \textit{Materials and
Methods} for further details). The first control measures were announced in
March 12, and the state of emergency was declared on March 18. We see that
after the lockdown the mobility of residents and non--residents decreased to
about 20\% and 45\% of the baseline. The maps show the spatial distribution of
the relative mobility for three specific days (March 9, March 19, and April
23), for residents (R1, R2, and R3) and for non--residents (N1, N2, and
N3).}\label{Mobility_Reduction}
\end{figure*}

\section{Introduction}
In response to the coronavirus pandemics, many governments have imposed
lockdowns, closing several public places, such as, schools, restaurants, sport
centers, and shopping malls~\cite{hale2020variation}. Companies and public
services have opted for giving employees the choice to work from
home~\cite{brynjolfsson2020covid}. These non-pharmacological interventions are
believed to be responsible for a significant slowdown of the spreading in
different countries
~\cite{maier2020effective,zhang2020changes,kraemer2020effect,buckee2020aggregated,remuzzi2020covid,giordano2020modelling,tian2020investigation,salje2020estimating}.
However, one still lacks an understanding on how the mobility patterns have
changed and how such changes affect the most probable pathways for the spread
of the virus.

With digitalization and advancements in global data collection, the public
expectations for knowledge-based strategies to control the spreading of the
coronavirus have been high. Daily, the public is swamped with statistics about
new infections, deaths, and recoveries around the world
~\cite{dong2020interactive,FinancialTimes2020}. Models and metrics are being
proposed and further developed to turn data into actionable insights
~\cite{y2018charting}. But, statistics of demographic variations of the impact
of a virus are not enough.  An extensive body of research on recent public
health threats, such as the 2013 MERS-CoV, 2014 Ebola, and 2016 Zika viruses,
has shown that to estimate the spread and implement efficient actions for
controlling it, one needs to rigorously investigate the mobility
patterns~\cite{halloran2014ebola,wesolowski2014commentary,poletto2014assessment,bogoch2016anticipating}.
Depending on the level of granularity, mobility data might be obtained from
civil aviation statistics~\cite{woolley2011complexity,barbosa2018human},
national censuses~\cite{barbosa2018human,gonzalez2008understanding}, public
transit
ridership~\cite{ponte2018traveling,caminha2017human,verma2020extracting}, or
mobile phone tracking \cite{tizzoni2014use,gonzalez2008understanding}.  By May
2020, there were more than $60$k travel restrictions issued around the world
to control inter- and intra-country travelling~\cite{DTM2020}. The mobility
patterns today are very different from the ones in February 2020 and these
publicly available data sets are no longer representative.

Epidemic spreading is a network-driven
process~\cite{pastor2015epidemic,gomez2018critical,meloni2011modeling,tizzoni2014use}.
To understand the complex spatiotemporal spreading dynamics, one needs to
replace the traditional view based on geographical distance by a
probabilistically motivated effective distance computed from the flux of
people~\cite{brockmann2013hidden}. With such a change of paradigm, one can
identify the set of most probable paths for the spreading of the virus and the
relative arrival times of an epidemics, which are both independent of the
epidemic parameters. Thus, the role of a municipality on the global dynamics
depends on its daily flow and effective distance to other municipalities.

Here, we study this impact by analyzing mobile phone data for the
inter-municipality daily flow of people in Portugal in March and April 2020.
The first confirmed case of COVID-19 in Portugal was reported on March 2,
2020. Ten days later, the total number of cases added up to 78 and the
government announced a first set of control measures, which included school
closures from March 14 onward. With a total of 642 identified cases, on March
19, the government decided to declare the state of emergency to impose more
drastic measures of mobility restriction and social distance. We show that,
the impact of the different measures was neither instantaneous nor homogeneous
throughout the country. It extended over several days and the relative impact
was strongly dependent on the population size of the municipality.  We also analyze
how the changes in mobility impacted on the distances. We show that
the effective distances changed differently for each municipality. The
structure of the set of most probable paths and relative arrival times also
changed, with a significant impact on the spreading dynamics.  

\begin{figure*}[t]
\includegraphics[width=\textwidth]{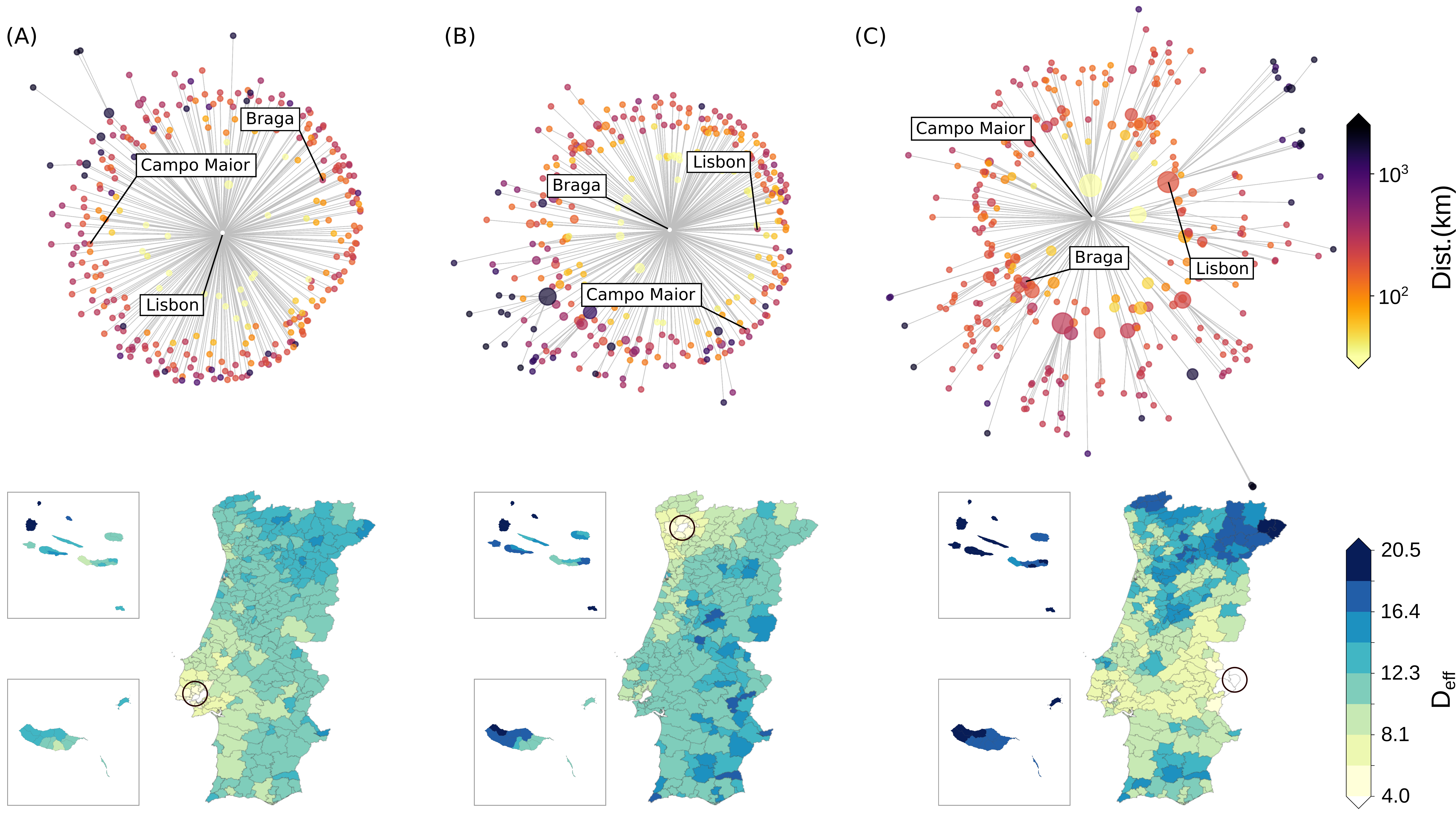}
\caption{\textbf{Shortest--path trees and effective distances before the
lockdown.} Shortest--path trees from Lisbon (A), Braga (B), and Campo Maior
(C) (central nodes). Each node is a different municipality and the distance to
the central node is proportional the effective distance $D_\mathrm{eff}$,
as defined in the \textit{Materials and Methods}. The effective distances are
calculated from the origin--destination matrix of the average mobility between
March 2 and 6. The nodes are colored according to the geographical distance to
the central node. In the bottom row are the country maps, where the color of
each municipality is given by the effective distance from the corresponding
central municipality marked by a black circle.~\label{Network}}
\end{figure*}

\begin{figure*}[t]
\includegraphics[width=\textwidth]{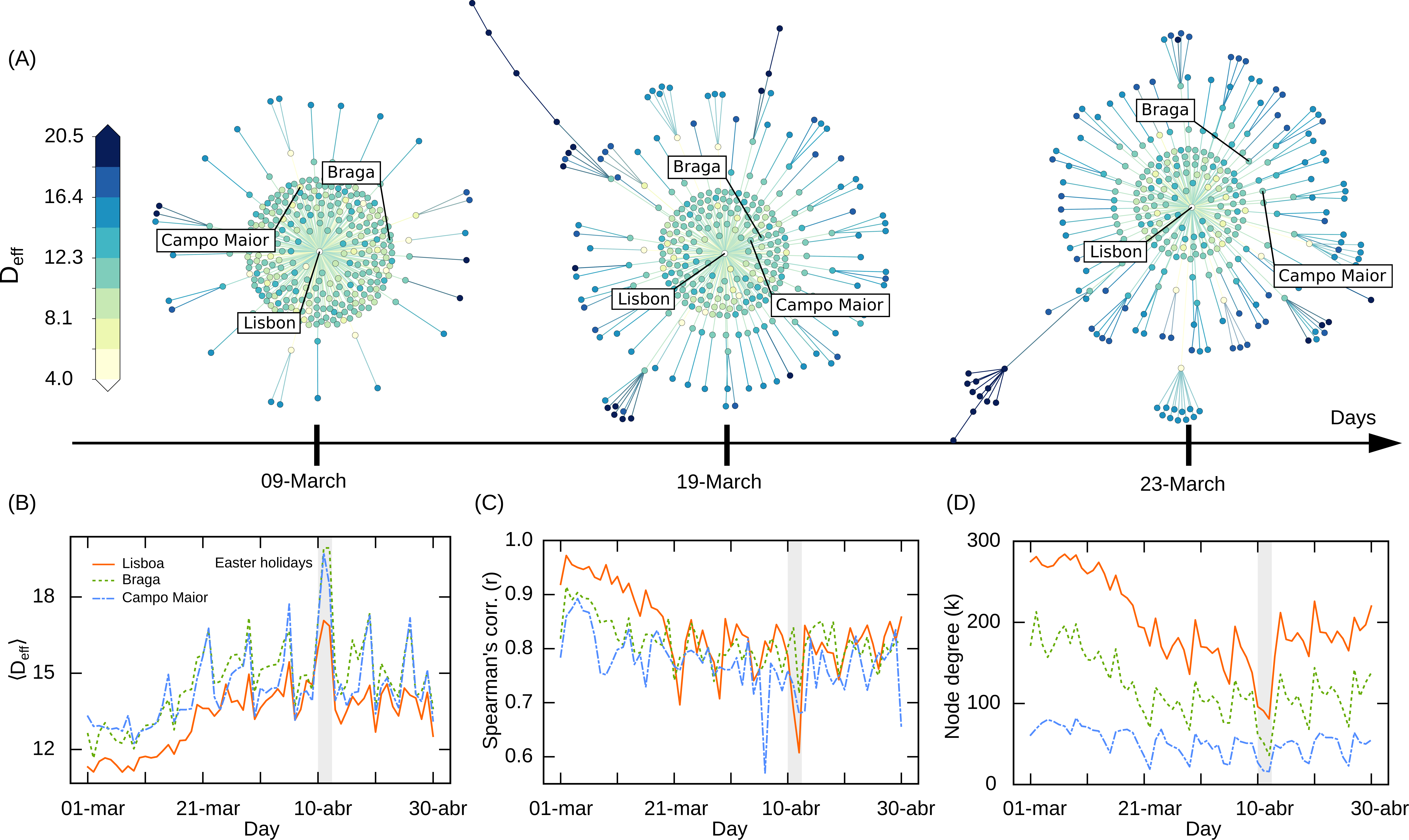}
\caption{\textbf{The effect of the lockdown on the mobility network.}  (A)
Shortest--path tree from Lisbon for three different days. The position of the
nodes is given by the Fain Hu layout~\cite{hu2005efficient}, designed to
distribute symmetrically the nodes and minimize edge crossing. The color of
each node is given by the effective distance from Lisbon. (B) Time dependence
of the average effective distance in the shortest--path trees shown in
Fig.~\ref{Network}. Before the lockdown, for these three municipalities,
Lisbon had the shortest average distance to the rest of the country, followed
by Braga and Campo Maior. With control measurements, the three average
effective distances increased, reaching a maximum in the Easter holidays, when
the Portuguese government imposed more severe mobility restrictions to
inter-municipality travels. (C) Spearman's correlation coefficient $r$ between
the effective distance from the average origin--destination matrix from March
2 to 6, and all the days of March and April.  With the lockdown, the rank of
the effective distances have changed as shown.  (C) Time evolution of the node
degree $k$ (number of edges) for Lisbon, Braga, and Campo Maior, for the
shortest--path trees shown in Fig.~\ref{Network}.  Initially, Lisbon was
connected directly to more than 90\% of the municipalities.  With the lockdown,
the $k$ decreases reaching a minimum at the Easter holidays, with only less than
50\% of the municipalities connected with Lisbon by only one
edge.}\label{NetworkChange}
\end{figure*}

\begin{figure}
\includegraphics[width=\columnwidth]{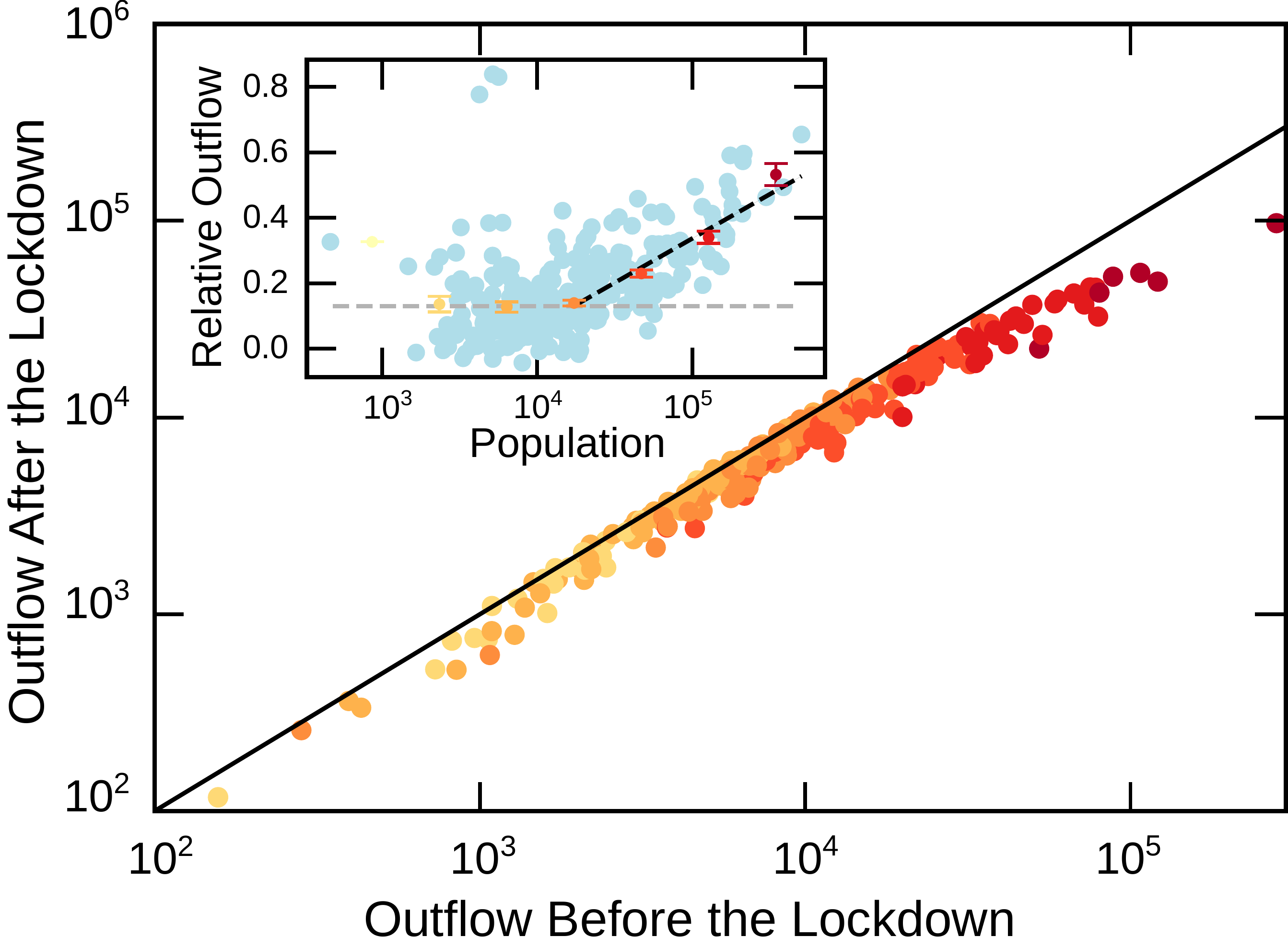}
\caption{\textbf{The impact of the lockdown depends on the mobility pattern of
the municipality.} Outflow of each municipality before and after the lockdown.
To reduce the statistical noise, the outflow corresponds to averages over five
days: before any control measurements, we considered March 2 to 6, and for
after the lockdown April 13 to 17. Each colored circle is a municipality and
the black solid line is a guide for the eyes and corresponds to a linear
relation between the two. The color of each data point represents the
population size of the municipality, following the log-spaced bins shown in
the inset. We can see that the effect of the lockdown is not homogeneous
across the country, municipalities with large values of mobility at the
beginning of March showed a large reduction in April. In the inset is the
relative change on outflow of each municipality (cyan circles) as a function
of its population size. The relative change is calculated by the outflow
difference between March and April divided by the one in March. The colors are
given by the average relative change on outflow in log-spaced bins (based on
the population size).  The dashed lines are best fits for a logarithmic
dependence of the relative mobility (RO) on the population size (P). For
smaller municipalities ($<18 000$ people), we find
$RO=0.034\log_{10}(P)-0.25$, while for the larger ones
$RO=-0.27\log_{10}(P)+1.02$. The population for each municipality was obtained
from Ref.~\cite{PorData}.}\label{Outflow}
\end{figure}

\section{Results}
We analyze a national-scale mobile dataset provided by NOS, a Portuguese
telecommunication company. This dataset contains the daily number of people
that moved within and between municipalities in Portugal in March and April
2020, obtained from the mobile recurrent usage of residents and non-residents
(see \textit{Materials and Methods} for further details).  The non-residents
are defined as users with a SIM card from a foreign mobile network. A user
is registered in the dataset if they stay for, at least, 60 minutes within
the boundaries of a municipality (see \textit{Materials and Methods} for
further details).

To quantify the inter-municipality mobility, we measure the daily number of
users moving within and between municipalities and define relative mobility as
this number rescaled by the average mobility from March 2 to 6 (baseline),
which is a period before any control measure.  In
Fig.~\ref{Mobility_Reduction}, we show the time dependence of the relative
mobility averaged over all municipalities for residents and non--resident
users. Even before the declaration of the state of emergency (March 18), with
the first set of control measures (March 12), the mobility started to
decrease. However, this change in mobility was not instantaneous, it occurred
over several days and only reached a steady level after March 21, with a
decrease of about $20\%$ for residents and $45\%$ for non--residents.  The
periodic structure observed in the steady regime corresponds to the expected
weekday (Monday to Friday) and weekend (Saturday to Sunday) differences, with
a clear decrease in mobility during the weekends. In the same figure, we show
also the spatial distribution of the relative mobility for three different
days (March 9, March 19, and April 23). One sees that, the impact of the
control measures was heterogeneous throughout the country. To characterize
these heterogeneities, in what follows, we analyze the inter--municipality
mobility.  Since only $3.3\%$ of the users are non-residents, for simplicity,
we consider only the resident one.

The origin--destination matrix (OD) is defined such that each element
$P_{ij}(d)$ is the fraction of users that, in day $d$, stayed for more than
$60$ minutes in both municipalities $i$ and $j$ (see \textit{Materials and
Methods}). From the OD matrix, we obtain the mobility network, which is a
weighted directed graph, where the nodes are the municipalities and the weight
of the link $ij$ is $P_{ij}$. Note that, in general, $P_{ij}\neq P_{ji}$, as
the total number of users in each municipality is different (see
\textit{Materials and Methods} for further details). The mobility network sets
the stage for the propagation to other municipalities. As shown in
Ref.~\cite{brockmann2013hidden}, further insight into the propagation dynamics
is obtained from the mobility network if we compute the effective distance
$D_\mathrm{eff}$ between nodes, defined as follows. For each pair of nodes $i$
and $j$, we defined the length of the link as $\ell_{ij}=1-\log(P_{ij})$.
Thus, the most probable path between two nodes $ij$ is the shortest path
between the two, defined as the one that minimizes the sum of all $\ell$ along
the path. The effective distance $D_\mathrm{eff}^{ij}$ is then the value of
such sum~\cite{brockmann2013hidden}. For every municipality $i$, we can obtain
the shortest--path tree, consisting of the set of the most probable paths to
all the other municipalities, where the central node is the origin $i$. In
Fig.~\ref{Network}, we show three examples of the shortest--path trees, for
(A) Lisbon, (B) Braga, and (C) Campo Maior, using an OD matrix obtained from
the average mobility from March 2 to 6. All nodes are placed at a distance
from the central node proportional to their effective distance to it. For the
sake of comparison, the color of a node is given by the geographic distance
from the central node. The size of the node is given by the value of the
betweenness centrality, which is defined as the number of shortest paths that
go from the central node to any other node on the tree through that node.  In
the figure, below each tree, we also show a color map displaying the spatial
distribution of effective distances from each center. It is clear that, some
municipalities can be geographically far from the central municipality, but
have a small value of effective distance, indicating that the relation between
effective distances and geographic distances are non-trivial.

Since the change in mobility was heterogeneous throughout the country, as
shown in Fig.~\ref{Mobility_Reduction}, it is expected that both the effective
distances and structure of the shortest--path trees have changed. To
illustrate such changes, we show in Fig.~\ref{NetworkChange}(A) the time
evolution of the shortest--path tree for Lisbon. We discover a change from
a star-like structure to a more branched structure. Figure~\ref{NetworkChange}(B)
shows the time evolution of the average effective distance $\left \langle
D_\mathrm{eff} \right \rangle$ to all other municipalities for Lisbon, Braga,
and Campo Maior. With the lockdown, $\left \langle D_\mathrm{eff}\right
\rangle$ increased, consistent with a much slower propagation dynamics. The
maximum on the Easter holidays (gray region) is reflective of a set of
additional measures imposed by the Portuguese government for those days to
avoid the traditional family gatherings, which included severe restrictions to
the inter-municipality mobility.  For each day $d$, we calculated the
Spearman's correlation between the effective distance
$D_\mathrm{eff}(d)$ and the initial effective distance
$\overline{D_\mathrm{eff}}$, obtained from the average mobility network from
March 2 to 6 (before the lockdown). The Spearman's correlation $r$ for the
entire set of municipalities is defined
as
\begin{equation}
r=\frac{\mathrm{cov}(\mathrm{rk_d},\mathrm{\overline{rk}})}{\sigma(\mathrm{rk_d})\sigma(\mathrm{\overline{rk}})}
, 
\end{equation}
where $\mathrm{rk_d}$ is the rank of $D_\mathrm{eff}(d)$,
$\mathrm{\overline{rk}}$ is the rank of $\overline{D_\mathrm{eff}}$,
$\mathrm{cov}(\mathrm{rk_d},\mathrm{\overline{rk}})$ is the covariance of the
rank, and $\sigma(\mathrm{rk})$ is the standard deviation. An $r=1.0$ means
that the rank of effective distances has not changed, while $r=0$ corresponds
to a new rank that is uncorrelated from the original one. The daily evolution of $r$ is shown in
Fig.~\ref{NetworkChange}(C). For the three municipalities, $r$ decreased with
the lockdown. This result suggests that the hierarchical organization of
municipalities in a spreading process was affected significantly by the
lockdown. 

Figure~\ref{NetworkChange}(D) depicts the evolution of the degree $k$ of the
central node (Lisbon, Braga, and Campo Maior), corresponding to the number of
municipalities that are directly connected to the central node in the
shortest--path tree. Before the lockdown, Lisbon was connected by a single
edge to $90\%$ of the municipalities. With the lockdown, Lisbon, in average,
became only connected directly to $55\%$ of the municipalities. This is in
line with a change from a star-like to a branched structure, as observed in
Fig.~\ref{NetworkChange}(A). We see a similar behavior for Braga, starting
with the fraction of municipalities connected directly, decreasing from $58\%$
to $33\%$, and for Campo Maior, where the drop is from $23\%$ to $15\%$. 

In Fig.~\ref{Outflow}, we compare the outflow before and after the lockdown.
The outflow is defined as the total number of resident users of a municipality
that were also identified in a different municipality during the day (see
\textit{Materials and Methods}). The values before the lockdown are averages
from March 2 to 6 and after the lockdown from April 13 to 17. We find that,
although the control measures were the same for the entire country, their
impact on the outflow strongly depends on the initial value of the outflow.
For municipalities with low values of outflow, we observe a linear relation
between the outflows before and after, which suggests that the relative impact
is the same.  However, for the municipalities with higher values of the
outflow, we observe a sublinear relation between the two. The larger the
initial value of the outflow the stronger is the impact of the control
measures. Since the outflow of a municipality correlates strongly with the
population size~\cite{simini2012universal}, in the inset of
Fig.~\ref{Outflow}, we plot the relative decrease in outflow as a function of
the population size. For municipalities with less than $18000$ users, the
outflow decreased by $10-20\%$, without a clear dependence on the population
size.  Whereas, for larger municipalities, the relative decrease in the
outflow scales logarithmically with the population size (dashed line) and it
exceeds $60\%$ for the largest municipality (Lisbon). 

\section{Conclusions}
Human mobility sets the stage for global spreading phenomena~\cite{gomez2018critical,guimera2005worldwide,lee2017morphology,barbosa2018human,brockmann2013hidden}.
Across the world, the most relevant non-pharmacological interventions to
contain the spreading of the coronavirus have been related to closing borders
and restricting the intra- and inter-country mobility, which has affected the
mobility patterns~\cite{tian2020investigation}.  Here, we studied the impact
of such mobility restrictions in the inter-municipality flow of people in
Portugal, in March and April of 2020. From a large dataset of mobile phone
users, we obtained the origin--destination matrix and the shortest--path trees
from all municipalities. These trees represent the most probable paths for the
spreading of the virus, when starting from a given
municipality~\cite{tizzoni2014use,brockmann2013hidden,piontti2014infection}.
We find that the relative decrease in mobility correlates with the population
size of a municipality. For small municipalities, with less than $18000$
mobile users, the outflow is decreased by about $10-20\%$, regardless the size
of the municipality. However, for larger municipalities, the relative decrease
in the outflow depend strongly on the population size and it even exceeds
$60\%$ for the largest municipalities. Such a heterogeneous change in mobility
affects deeply the structure of the shortest--path trees and, therefore, the
relative effective distance between municipalities. 

To circumvent the absence of accurate data for mobility, models have been
developed to artificially generate them, as for example the gravity and
radiation models~\cite{de2011modelling,simini2012universal}, or to account for
mobility restrictions in the
models~\cite{arenas2020mathematical,costa2020metapopulation,meloni2011modeling}.
From our findings, it is now possible to extend these models to account for
the dependence on the size and mobility of a municipality. This heterogeneous
response is an important ingredient to consider in epidemic models, such as
the ones developed in
Refs.~\cite{arenas2020mathematical,costa2020metapopulation,meloni2011modeling,wang2014spatial,hollingsworth2006will,cooper2006delaying,bajardi2011human,ferguson2006strategies,colizza2007modeling,wang2012estimating},
so that governments can respond to the spread more accurately.

\section{Materials and Methods}

\subsection{Mobility matrix from mobile phones records}

The mobile data was obtained from NOS cellular network and its volume goes up
to about 8 Terabytes of events, collected from the network interfaces in March
and April 2020.  Each record in the dataset refers to a unique subscriber id,
General Data Protection Regulation (GDPR) compliant regarding Personal
Identifiable Information (PII)~\cite{GDPR}, and it includes the date/timestamp
of each respective event and the geographic coordinates of the respective
network cell coverage centroid where each event occurred. Such dataset can
thus provide an abstraction of subscriber physical displacements over time.

The process to compute the mobility matrices is composed of several steps:
from data event aggregation and location estimation to trip count aggregation.
Within the data event aggregation part, the events are aggregated in a $10$
minutes sliding window with detection counts for each subscriber id performed
by cell. The location for each subscriber id is estimated based on the most
frequent cell algorithm. Once the location trips are extracted, the process to
derive origin--destination flows between geographic areas is the following: i.
The geographic area under analysis is divided into municipalities; ii. Origin
and destination municipalities, together with starting time are extracted for
each subscriber id.  iii. Trips with the same origin, destination, and time
frame are grouped together. The result is a matrix whose element represents
the number of trips from origin $i$ to destination $j$, starting within the
time frame.  Here, we used the municipal areas in Portugal for the regions and
the time frame of 6 am and 8 pm.

From the mobility matrix, the extrapolation process is carried out to
represent the whole universe of the population in study. The extrapolation
algorithm takes into consideration local as well as international market
shares, besides wholesale agreements between parties.  After extrapolation,
the transformation process takes place to extract features from
\emph{residents} and \emph{non--residents}, segmenting the data between flows
from region $i$ to $j$, and the ones that do not move within that time window,
which is the same as saying that $i$ is equal to $j$.  At the end, compliance
with the data privacy rules (GDPR) are applied and trips with counts under 6
are omitted.

\subsection{Shortest--path trees}

For the network analysis, we only considered the movement of residents, i.e.,
in the mobility matrix $\textbf{F(d)}$, $F_{ij}(d)$ is the number of resident
users who were identified at both municipality $i$ and $j$ within the day $d$. In
order to measure the combined number of individuals from $i$ to $j$, we
considered the undirected mobility network using the standard symmetrization operation
$\textbf{F}=\textbf{F}+\textbf{F}^T$, where $\textbf{F}^T$ is the transposed
matrix of $\textbf{F}$~\cite{brockmann2013hidden}. 

Each element of the origin--destination matrix $\textbf{P}$ is calculated as
$P_{ij}=F_{ij}/\sum_{n} F_{in}$, where the sum is over all municipalities, and
quantifies the mobility network, where the nodes are the municipalities and
the edges between $i$ and $j$ have a weight $P_{ij}$.  If we define the
distance $\ell_{ij}=1-\log(P_{ij})$ for each edge, the most probable paths are
the ones that minimize the total sum of $\ell$~\cite{brockmann2013hidden}. For
each node, one can define a shortest--path tree (see Fig.~\ref{Network}) and
the effective distance $D_\mathrm{eff}$ from a node $i$ to $j$ as the length
of the shortest path between these two nodes. 

\begin{acknowledgments}
We acknowledge financial support from the Portuguese Foundation for Science
and Technology (FCT) under Contracts no. PTDC/FIS-MAC/28146/2017
(LISBOA-01-0145-FEDER-028146), UIDB/00618/2020 and UIDP/00618/2020, and from
NOS SGPS S.A. for providing the mobile data.
\end{acknowledgments}

\bibliography{paper.bib}

\end{document}